\documentclass[prl,twocolumn,showpacs,superscriptaddress]{revtex4}

\usepackage{graphicx}
\usepackage{amsmath}
\usepackage{amssymb}
\usepackage{epsfig}
\usepackage{subfigure}

\renewcommand{\v}[1]{{\bf #1}}

\newcommand{\w}{{\omega}}

\def\eqa{\begin{eqnarray}}
\def\eea{\end{eqnarray}}
\newcommand{\eq}{\begin{equation}}
\newcommand{\ee}{\end{equation}}
\newcommand{\nn}{\nonumber\\}

\newcommand{\<}{\langle}
\renewcommand{\>}{\rangle}
\newcommand{\Tr}{{\rm Tr}}

\renewcommand{\Im}{{\rm Im}}

\newcommand{\ra}{\rightarrow}

\newcommand{\del}{\delta}
\newcommand{\Del}{\Delta}

\renewcommand{\th}{\theta}

\newcommand{\si}{\sigma}

\usepackage{color}
\begin{document}

\title{Visualizing the $d$-vector in a nematic triplet superconductor}

\author{Wei-Cheng Bao}
\affiliation{National Laboratory of Solid State Microstructures $\&$ School of Physics, Nanjing
University, Nanjing, 210093, China}
\affiliation{Zhejiang University of Water Resources and Electric Power, Hangzhou 310018, China}

\author{Qing-Kun Tang}
\affiliation{National Laboratory of Solid State Microstructures $\&$ School of Physics, Nanjing
	University, Nanjing, 210093, China}

\author{Da-Chuan Lu}
\affiliation{National Laboratory of Solid State Microstructures $\&$ School of Physics, Nanjing
	University, Nanjing, 210093, China}

\author{Qiang-Hua Wang}
\email{qhwang@nju.edu.cn}
\affiliation{National Laboratory of Solid State Microstructures $\&$ School of Physics, Nanjing
University, Nanjing, 210093, China}
\affiliation{Collaborative Innovation Center of Advanced Microstructures, Nanjing 210093, China}

%\date{\today}

\begin{abstract}
Recent experiments show strong evidences of nematic triplet superconductivity in doped Bi$_2$Se$_3$ and in Bi$_2$Te$_3$ thin film on a superconducting substrate, but with varying identifications of the direction of the $d$-vector of the triplet that is essential to the topology of the underlying superconductivity. Here we show that the $d$-vector can be directly visualized by scanning tunneling measurements: At subgap energies the $d$-vector is along the leading peak wave-vector in the quasi-particle-interference pattern for potential impurities, and counter-intuitively along the elongation of the local density-of-state profile of the vortex. The results provide a useful guide to experiments, the result of which would in turn pose a stringent constraint on the pairing symmetry.

\end{abstract}

%
%75.30.Fv  Spin-density waves
%74.20.Rp  Pairing symmetries (other than s-wave)
%74.20.-z  Theories and models of superconducting state
%71.27.+a  Strongly correlated electron systems; heavy fermions
%64.60.ae  Renormalization-group theory

\maketitle

{\em Introduction}: Topological superconductors (TSC's) have been attracting extensive interest because of the potential application in topological quantum computing. \cite{Nayak} One type of such superconductors are time-reversal-breaking (TRB) $p+ip'$-wave superconductors. In the spinless (or spin polarized) case, of topological class D, which may be realized in the $\nu=5/2$ fractional quantum Hall system, each vortex hosts a Majorana zero mode (MZM), and braiding vortices leads to permutation of the degenerate many-body states,\cite{Read,Ivanov} enabling non-Abelian statistics that can be utilized in fault-tolerant quantum computation. In the spinful case (of topological class A), which is possibly realized in Sr$_2$RuO$_4$,\cite{Maeno} the usual vortex of flux quantum $hc/2e$ hosts a canonical Fermion zero mode instead. However, single MZM can be realized in a vortex of half flux quantum, $hc/4e$.\cite{Ivanov} Strong evidence of such a half-quantum vortex has been observed in a recent experiment.\cite{Jang} The other type of TSC's, of topological class DIII, are helical or time-reversal invariant (TRI). In this case MZM's come in pair in each vortex. The braiding of such vortices in two-dimension (2d) or 3d is not yet fully understood. Interestingly, however, in a 1d chain pair of MZM's appear on the ends, and it has been shown that non-Abelian statistics can still be realized by braiding such pairs in a T-junction.\cite{Law} 

There are strong evidences of DIII-TSC in some doped Bi$_2$Se$_3$ crystals. The undoped Bi$_2$Se$_3$ is a TRI topological insulator (TI). The strong spin-orbital coupling in TI makes DIII-TSC likely, although the topology of TI is not necessary.\cite{Fu-Berg} Doped Cu$_x$Bi$_2$Se$_3$ is found to be superconducting,\cite{Cava,Ando} and early specific heat measurement \cite{Ando} seems to indicate a full pairing gap. In this material, the superconducting state would be topological if it is fully gapped and the pairing function is odd under inversion.\cite{Fu-Berg} The pairing function is initially proposed as $\si_2 \v V\cdot \v s$ with a $d$-vector $\v V=\hat{z}$, dubbed $\Del_2$ pairing. \cite{Fu-Berg} Henceforth $\si_{1,2,3}$ and $\v s=(s_1, s_2, s_3)$ are Pauli matrices acting on (two effective) orbitals and (pseudo-) spins. However, the subsequent point-contact tunneling measurement reveals a zero-bias peak (ZBP) in the spectrum,\cite{ZBP} which could be interpreted either as a result of nodal pairing gap, or as a smearing up of density of states from the material-dependent twisted surface bands of the above fully gapped TSC.\cite{twist} In contrast, the scanning-tunneling-microscopy (STM) measurement on the same $(1,1,1)$ surface of the same material reveals a full gap, with no sign of in-gap surface states expected of TSC.\cite{Levy} Recently, it is found that the Knight shift develops two-fold oscillation for in-plane applied fields, with strongest (or no) suppression below $T_c$ for $H$ along (or orthogonal to) one of the Se-Se bonds, the $x$-direction henceforth, suggesting a triplet Cooper-pair with its $d$-vector along $x$. \cite{NMR} Fu realized that the inplane nematicity in the Knight shift implies the pairing function must be in a doublet representation of the symmetry group.\cite{Fu} Indeed, the two cases of $\si_2\v V\cdot \v s$, with $\v V=\hat{x}$ (dubbed $\Del_{4x}$) and $\v V=\hat{y}$ (dubbed $\Del_{4y}$) respectively, form a doublet $E_u$ representation, transforming as $x$ and $y$ under $C_3$. In the continuum limit of the $\v k\cdot\v p$ theory, these pairing functions develop anti-nodal (or nodal) quasiparticle gap at Fermi momenta along (or orthogonal) to the $d$-vector, dooming the full gap required for TSC. However, the warping term allowed in the lattice can help remove the node in the $\Del_{4y}$ case,\cite{Fu} although $\Del_{4x}$ remains nodal since the warping term vanishes on mirror planes. Therefore, only $\Del_{4y}$ is topologically nontrivial in a strict sense. 
   
The nematicity in the superconducting states are also observed in angle-resolved specific heat,\cite{Yenozawa} upper critical field \cite{Yenozawa, Pan, Smylie} and magnetic torque \cite{Asaba}, but with varying identifications of the $d$-vector. The specific heat observed in Cu$_x$Bi$_2$Se$_3$ is smallest (or largest) for field applied along (orthogonal to) $x$,\cite{Yenozawa} implying $\Del_{4y}$ pairing with gap-minimum along $x$. In the same experiment,\cite{Yenozawa} it is observed that the in-plane upper critical field $H_{c2}$ is largest along $x$ (orthogonal to the $d$-vector). However, the $d$-vector is identified as along the direction where $H_{c2}$ is maximal in Sr$_x$Bi$_2$Se$_3$.\cite{Smylie} Notice that $\Del_{4x}$ and $\Del_{4y}$ may coexist in vortices near $H_{c2}$, leading to material-dependent anisotropy.\cite{GL} 

It is therefore important to identify the $d$-vector of the possible nematic triplet superconductor. Here we show that apart from the experimental probes mentioned above, it is possible to visualize the $d$-vector directly by STM: At subgap energies the $d$-vector is along the leading peak momentum in the quasi-particle-interference pattern (QPI) in the presence of a potential impurity, and is (counter-intuitively) along the elongation of the local density-of-state profile (LDOSP) of the vortex (to be distinguished from the OPP). Our theory is based on a 2d lattice model, which we hope is reasonable for doped Bi$_2$Se$_3$, given the weak Van der Waals coupling between the quintuple layers. We notice that superconductivity is reported to be induced in Bi$_2$Se$_3$ on superconducting substrates.\cite{proximity, Jia} More recently, nematicity is clearly observed in a thin film of Bi$_2$Te$_3$ on a FeTe$_x$Se$_{1-x}$ superconductor.\cite{Wen} Our results should apply directly in the latter cases when the $E_u$ triplet is present.

\begin{figure}  
	\includegraphics[width=0.95\columnwidth]{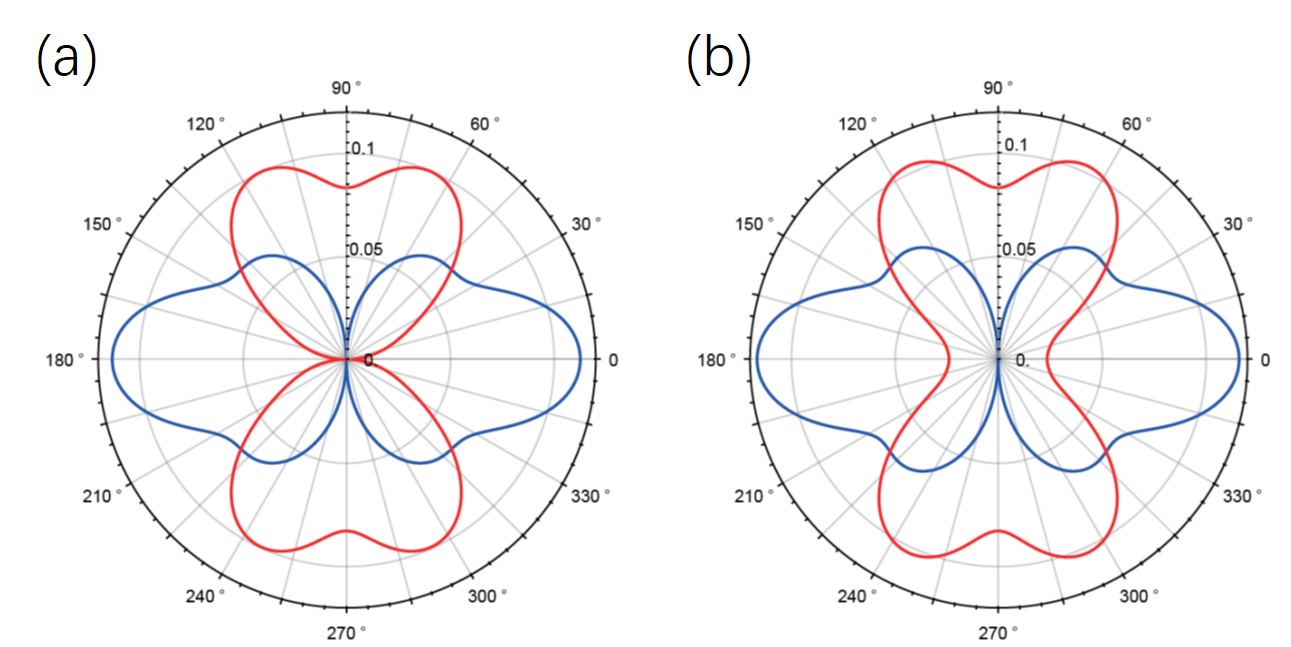}
	\caption{Quasiparticle gap, defined as the minimal excitation energy along a given polar direction, versus the polar angle. Here we set $\Del_0=0.26$, but the resulting characteristic gap scale is $\Del=0.1$. The blue (red) line corresponds to $\v V=\hat{x}$ ($\v V=\hat{y}$). The warping parameter is $t_{\rm w}=0$ in (a) and $t_{\rm w}=0.2$ in (b).}
	\label{fig:gap} %% label for entire figure
\end{figure}

{\em Model}: Following the continuum model described in Ref.\cite{Fu-Berg,Fu}, we define a 2d triangular lattice model for a doped TI in order to be more specific to the material. The normal-state part of the Hamiltonian is given by, in the momentum space, $H_0 = \sum_\v k \psi_\v k^\dag h_\v k\psi_\v k$, with $\psi_\v k$ a four-component spinor describing two $p_z$-like orbitals from Se atoms (related by inversion) and two pseudo-spins (spins henceforth for brevity), and 
\eqa h_\v k =&&\sum_i (\hat{z}\times \v d_i)\cdot \v s \si_3 \sin k_i +t_{\rm w}\sum_i \si_3 s_3 f_i \sin k_i \nn
&&+(m+3- \sum_i\cos k_i )\si_1 -\mu.  
\eea
Here $\v d_1=(1,0,0)$, $\v d_2=(1/2,\sqrt{3}/2,0)$ and $\v d_3=(-1/2,\sqrt{3}/2,0)$ are three translation vectors, $k_i=\v k\cdot \v d_i$, $t_{\rm w}$ is the warping parameter, $f_{1,2,3}=(1,-1,1)$ is an $f$-wave form factor, $m$ controls the topology of the normal state band structure, and $\mu$ is the chemical potential. Throughout this work we use arbitrary units for qualitative purposes. For concreteness we choose $m=-0.5$, $\mu=3.25$, and $t_{\rm w}=0.2$, unless specified otherwise. Notice that under inversion $\v k\ra -\v k$, $\v s\ra\v s$ and ${\bf \si} \ra\si_1 {\bf \si}\si_1$, and under TR, $\v k\ra -\v k$, $\v s\ra -\v s$ and ${\bf \si}\ra {\bf \si}^*$. The Hamiltonian $H_0$ is explicitly TRI and inversion symmetric, and describes a TI if $\mu=0$. The effective Hamiltonian in the superconducting state is, in the Nambu basis $\Psi_\v k^\dag = (\psi_\v k^\dag, -i\psi_{-\v k}^t s_2)$,
\eqa H = \frac{1}{2}\sum_\v k \Psi_\v k^\dag M_\v k \Psi_\v k, \ \ \ M_\v k = h_\v k \tau_3 + \Del_0 \v V\cdot\v s\si_2\tau_1,\eea
where the factor of $1/2$ reduces double counting due to the Nambu basis, $\Del_0$ is the macroscopic order parameter (assumed real for the moment), $\v V$ is the uni-modulus $d$-vector of the triplet. Henceforth the Pauli matrices $\tau_{1,2,3}$ act on Nambu basis. The quasiparticles are described by the eigenstates of $M_\v k$. The resulting quasiparticle gap depends on the direction of $\v V$ and the warping parameter $t_{\rm w}$, as shown in Fig.\ref{fig:gap}. For $\Del_0=0.26$ the characteristic gap size is $\Del\sim 0.1$. In agreement with the continuum model,\cite{Fu} it is nodal along the $y$-direction (where the warping function vanishes) for $\v V=\hat{x}$ [see (a) and (b)], while it is nodal along $x$ for $\v V=\hat{y}$ (a) unless the warping  parameter $t_{\rm w}$ is finite (b). Notice that similar gap structure appears for $\v V$ along other directions related to the above ones by $C_6$ rotations. 

\begin{figure}
	\includegraphics[width=0.95\columnwidth]{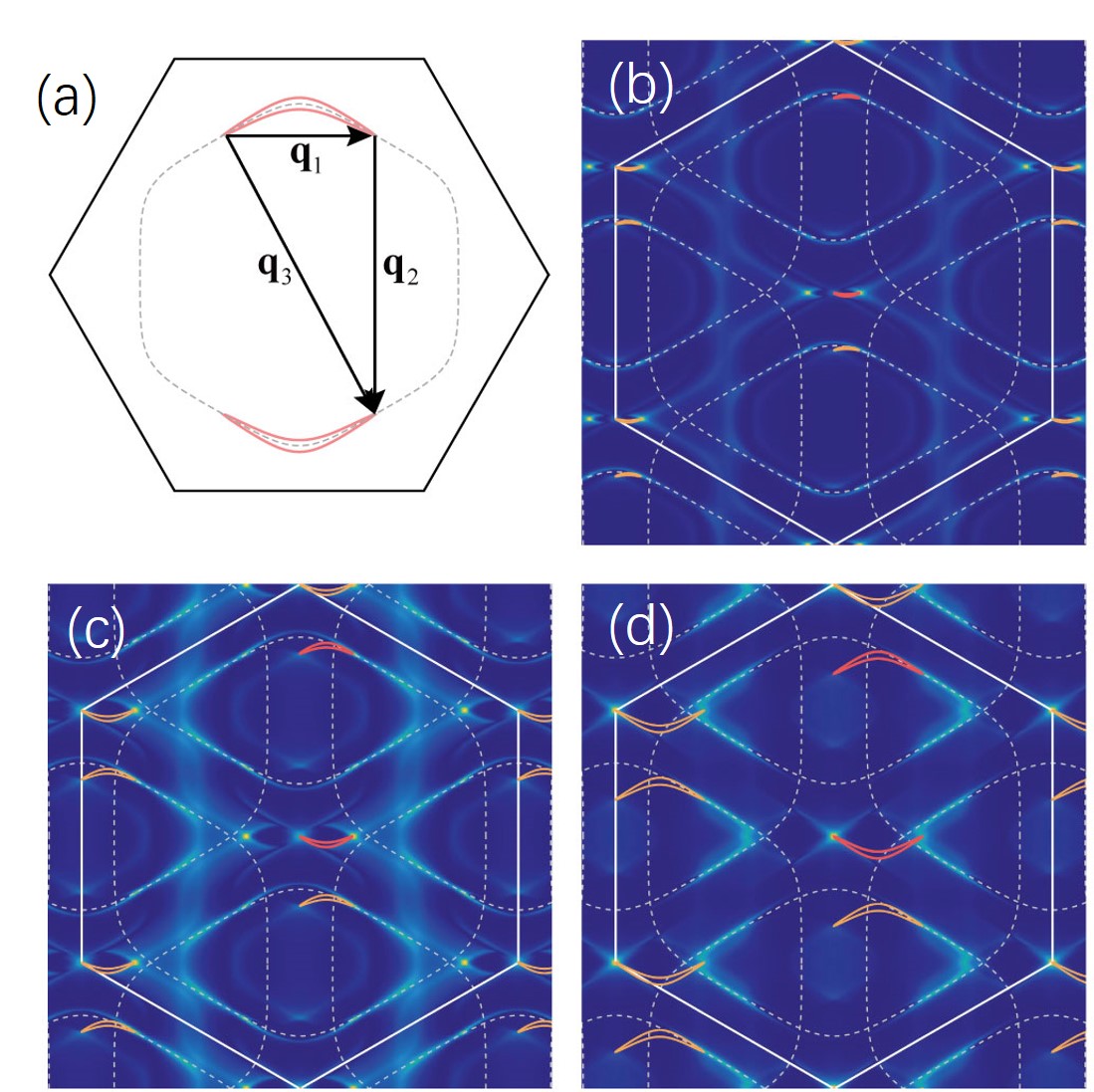}
	\caption{QPI for $\v V=\hat{x}$. (a) Typical EEC at a low energy, together with leading scattering vectors (aside from symmetry-related ones). The QPI spectrum $p(\v q,\w)$ (color scale) is shown at $\w=0.2\Del$, $0.4\Del$ and $0.6\Del$ in (b)-(d), respectively. The signal is higher where the color is brighter. Here we overlay the EEC's (red) in the reduced Brillouin zone, with one of the tips set at the origin. The EEC's (orange) in the extended zone, related to those in the reduced zone by reciprocal vectors (sides of the white hexagon), are also shown. The dashed lines are the $2k_f$-contours, duplicated by the reciprocal vectors. In this way, the importance of the scattering vectors connecting the origin to the tips of all EEC's, and of the $2k_f$-scattering, can be directly visualized.}
	\label{fig:qpi-x}
\end{figure}

\begin{figure}
	\includegraphics[width=0.95\columnwidth]{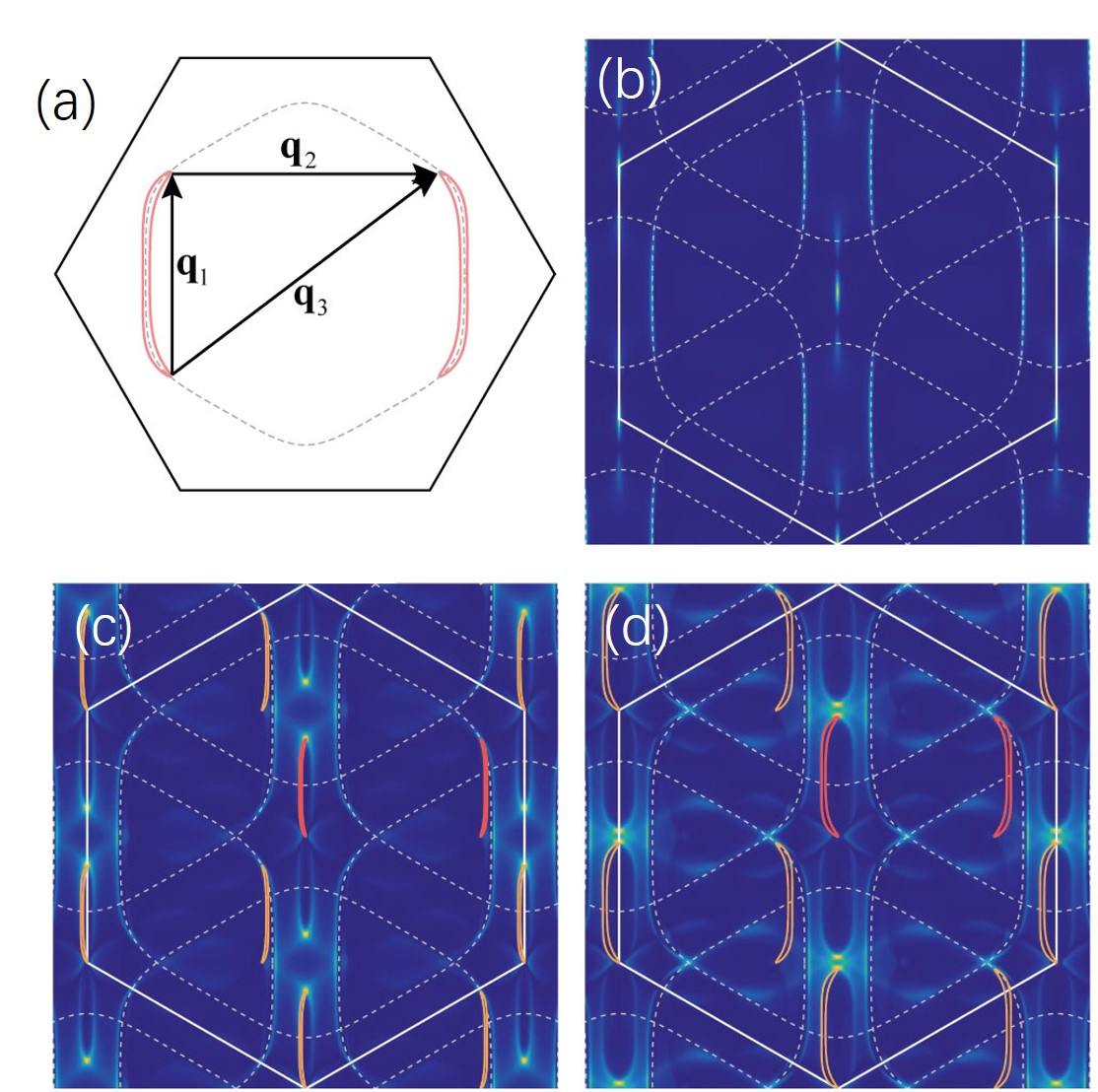}
	\caption{The same plot as Fig.\ref{fig:qpi-x} but for $\v V=\hat{y}$. The energy is $\w=0.2\Del$, $0.4\Del$ and $0.6\Del$ in (b)-(d), respectively.}
	\label{fig:qpi-y}
\end{figure}

{\em QPI}: The LDOS is uniform in a translationally invariant system. In the presence of an impurity, however, elastic scattering between equal-energy quasiparticle states leads to standing waves, or energy dependent ripples in LDOS. This effect is called QPI.\cite{Wang} The characteristic wave-vector of the ripples depends on the gap structure, and for this reason QPI has been applied with great success in the context of cuprate and iron-based superconductors. Here we show it is equally powerful to reveal the gap structure, the $d$-vector in particular, in a nematic triplet superconductor described above. 
We add to $H$ a scalar potential impurity at the origin, $H_{\rm imp}= \Psi_0^\dag V_{\rm s}\tau_3 \Psi_0$ at the real-space origin. Other types of impurities are less revealing (possibly due to matrix-element effect) and will be discussed in Supplementary Materials (SM). We set $V_{\rm s}=0.5$ for illustration. The variation in the LDOS (on top of the uniform background) at energy $\w$ and spatial position $\v r$ in the presence of the impurity is obtained as
\eqa \del\rho(\v r,\w) = -\frac{1}{\pi}\Im \left[\Tr P G_0(\v r, \w)T(\w) G_0(-\v r,\w)\right],\eea 
where $P=(\tau_0+\tau_3)/2$ is a projection operator, $G_0(\v r, \w)$ is the unperturbed and retarded Green's function, and $T$ is a matrix given by $T^{-1}(\w) = V_{\rm s}^{-1}\tau_3 - G_0(0,\w)$. More details can be found in SM. From $\del\rho(\v r,\w)$ we obtain the power spectrum 
$p(\v q,\w) = |\del\rho(\v q,\w)|^2$, where $\del\rho(\v q,\w)$ is the Fourier transform of $\del\rho(\v r,\w)$. In Fig.\ref{fig:qpi-x} we show the spectrum for $\v V=\hat{x}$ at selected ingap energies. Panel (a) shows the typical equal-energy contour (EEC) of quasiparticles near the Fermi surface, together with three leading elastic scattering vectors $\v q_{1,2,3}$ (aside from symmetry-related ones) with highest joint DOS. At energy $\w=0.2\Del$ (b), the peak momentum is $\v q_1$, the scattering vector connecting the two tips of a closed EEC. The reason that the $\v q_1$-scattering at low energies is stronger can be attributed to the fact that the low energy states are essentially normal Bloch states with spin-orbital-momentum texture. The matrix element of the scattering potential is larger in forward-like scattering processes (with shorter $\v q$). The vector $\v q_1$ increases in length as $\w$ increases (c and d). At the same time there are weaker features at $\v q_2$ and $\v q_3$ (which is also a $2k_f$-vector), and features on part of the $2k_f$-line. In addition, there are other features that can not be understood by the simple picture of on-shell scattering. (Off-shell scattering do contributes but is more difficult to understand.) Moreover, in the lattice theory $p(\v q+\v G,\w)=p(\v q,\w)$, where $\v G$ is a reciprocal vector, after the FT of $\del\rho(\v r,\w)$. This causes complicated replicating of the images, specific to the trigonal lattice. The replication could be removed by performing FT of interpolated $\del\rho(\v r,\w)$ on a denser grid, mimicking the experimental STM measurements. See SM for further details.

Fig.\ref{fig:qpi-y} is the same plot as Fig.\ref{fig:qpi-x} but for $\v V=\hat{y}$. We see bright features only at the origin and part of the $2k_f$-line at $\w=0.2\Del$. These are not from on-shell scattering since the energy is still below the minimal gap (hence no EEC's are drawn). Clear dispersive features develop at larger energies and at $\v q_1$ (modulo $\v G$) along $y$, or the direction of $\v V$, as seen in (b)-(d). The features at $\v q_{2,3}$ and on $2k_f$-line are much weaker. Taking Figs.\ref{fig:qpi-x} and \ref{fig:qpi-y} together, we conclude that the low energy QPI is dominated by $\v q_1$-scattering exactly along the $d$-vector $\v V$ of the nematic triplet.

\begin{figure}
	\includegraphics[width=\columnwidth]{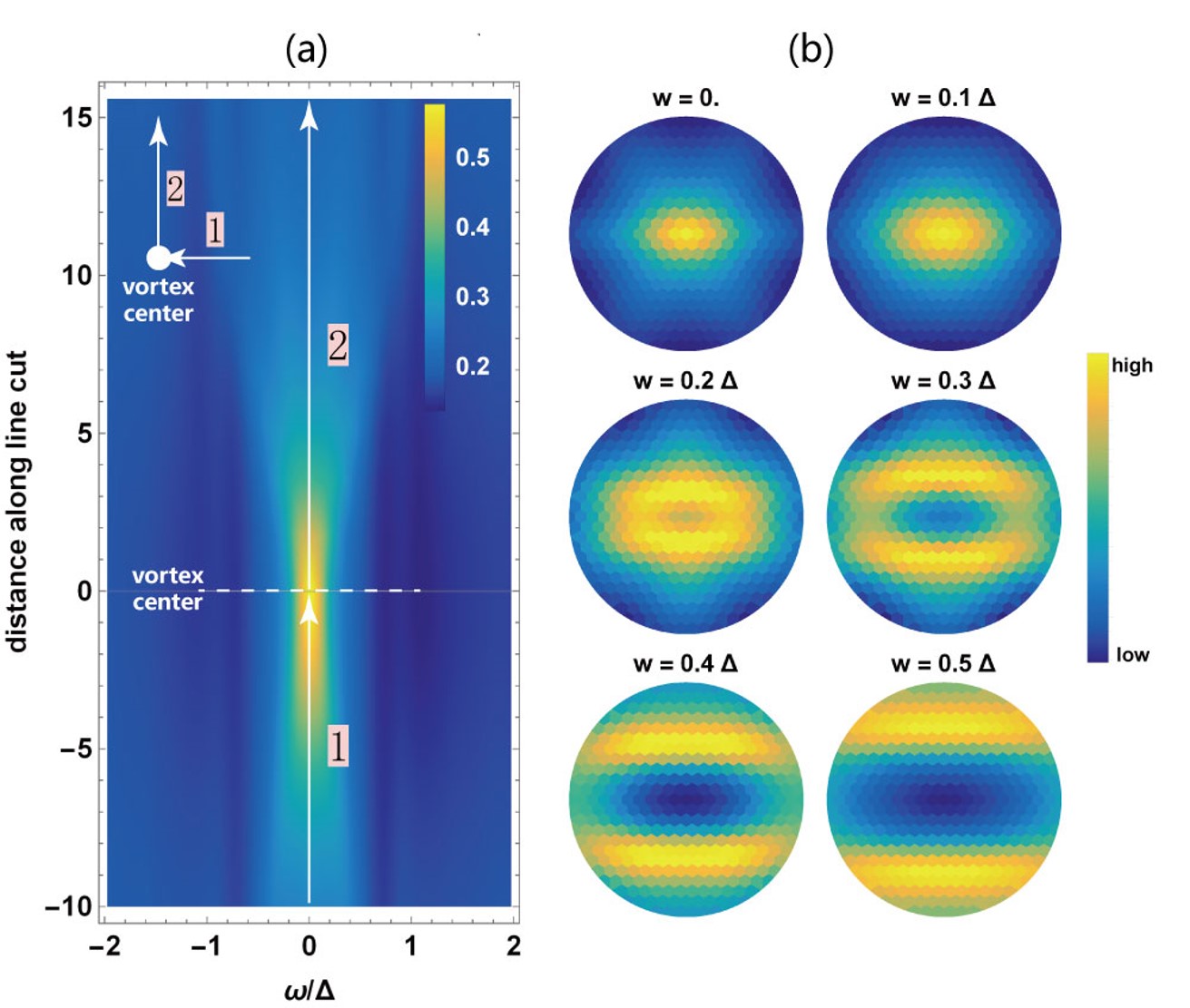}
	\caption{LDOS along the line-cut (a) and vortex profile at selected energies (b). Here $\v V=\hat{x}$.}
	\label{fig:vortex_Vx}
\end{figure}

\begin{figure}
	\includegraphics[width=\columnwidth]{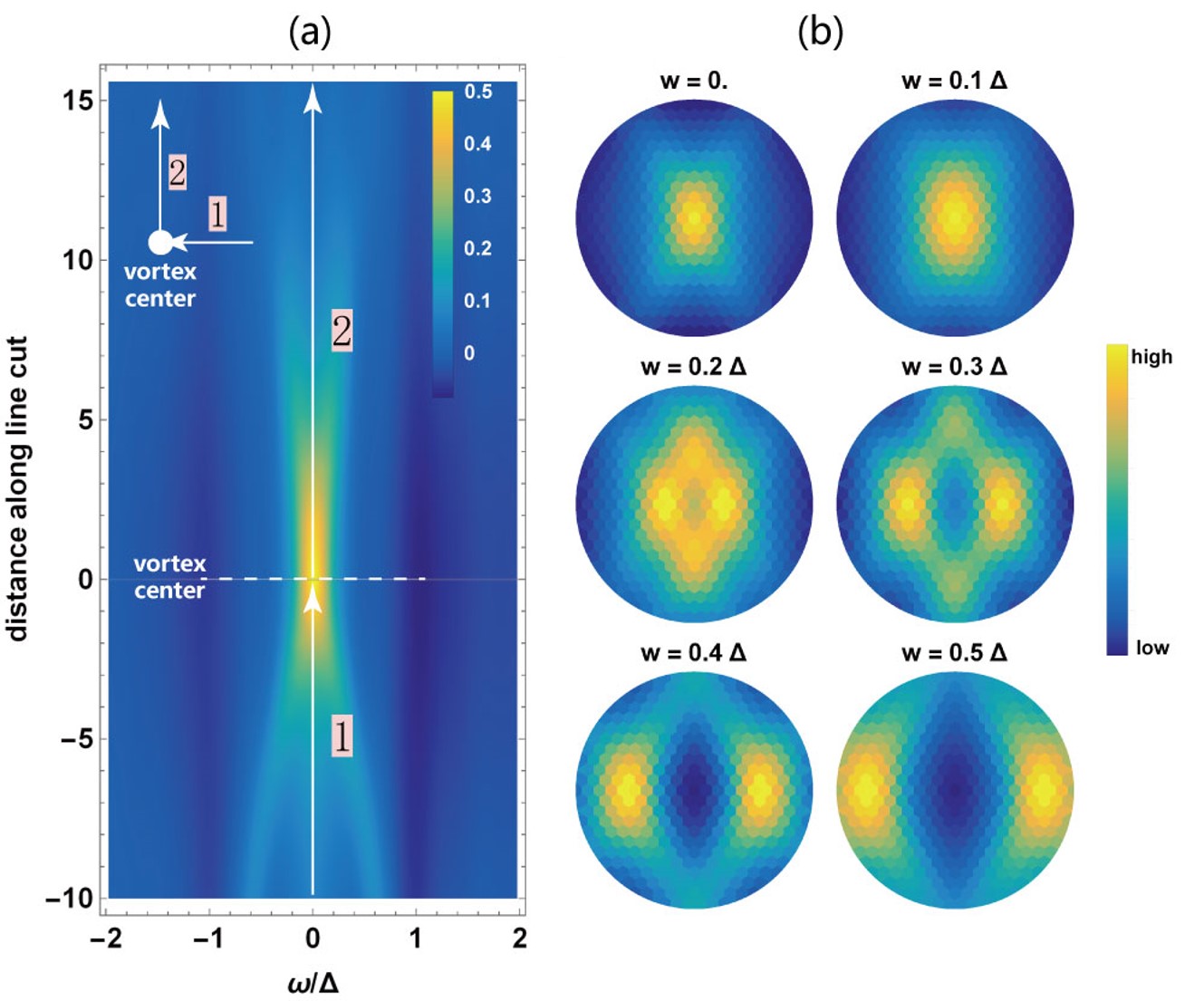}
	\caption{LDOS along the line-cut (left panels) and vortex profile at selected energies (right panels). Here $\v V =\hat{y}$.}
	\label{fig:vortex_Vy}
\end{figure}

{\em Vortex}: We now consider a vortex in an open hexagon. To gain sufficient resolution in energy, we consider a large-scale open hexagon with $101$ sites on each edge. We need to rewrite the Hamiltonian in real space, 
\eqa H = \frac{1}{2}\sum_{\v r,\v r'}\Psi^\dag_\v r M(\v r, \v r')\Psi_{\v r'}.\eea 
In the uniform state $M(\v r, \v r')=M_0(\v r,\v r')$ is just the Fourier transform of $M_\v k$. In the presence of a vortex, the normal part of $M$ is changed into $M_0(\v r, \v r') e^{i\int_{\v r'}^\v r \v A\cdot d\v l~\tau_3}$,  
and the anomalous (pairing) part becomes $\del_{\v r\v r'}\Del_0(\v r) e^{i\th} \v V\cdot \v s \si_2 \tau^+ + {\rm h.c.}$.
Here $\tau^+=(\tau_1+i\tau_2)/2$, $\v A = (0, Bx, 0)$ is the vector potential for a magnetic field $B\hat{z}$ in the Landau gauge, $\Del_0(\v r)$ is the order parameter profile (OPP) and $\th$ is the azimuthal angle of $\v r$, and we assume the vortex core is at the origin.  We use the units such that $\hbar=c=e=1$, and we assume $B$ is uniform within the system (applicable in the extreme London limit). We thread the system by a single flux quantum ($hc/2e$) so that $B=\pi/S$, where $S$ is the area. The following results for a single vortex are not sensitive to the details of $B$, and are in fact not changed qualitatively even if we set $B=0$. We consider $\v V=\hat{x}$ and $\v V=\hat{y}$ separately. We set an isotropic OPP, $\Del_0(\v r)=\Del_0 \tanh(r/\xi)$ where $\xi$ is representative of the coherence length. We set $\xi=10$ for illustration. (In the SM we discuss the effect of anisotropy in OPP up to a factor of $10$, which does not change our results qualitatively.) The LDOS is given by
\eqa \rho(\v r,\w) = \sum_{\nu = 1}^4 \<\v r\nu|\del(\w - M)|\v r\nu\>,\eea
where $|\v r\nu\>$ denotes a single-particle state localized at position $\v r$ with its spin-orbital basis labeled by $\nu$. The delta-function is resolved in terms of Chebyshev polynomials (see SM for details).
This enables us to calculate the LDOS on selected sites independently and most efficiently.  

Fig.\ref{fig:vortex_Vx}(a) shows the LDOS along two linecuts (inset). There is a clear ZBP near the vortex core, and this peak extends away from the vortex, followed by splitting into two peaks. The extension of the ZBP along $x$ (cut 1) is apparently much longer than that along $y$ (cut 2), and is therefore very anisotropic. To have a better view of the nematic anisotropy, we show in Fig.\ref{fig:vortex_Vx}(b) LDOS maps, or LDOSP's, in a circular view field near the vortex core, at selected energies. All these maps show elongation along $x$, the direction of the $d$-vector $\v V$. Fig.\ref{fig:vortex_Vy} shows the case of $\v V=\hat{y}$. Here ZBP is also present (a), but it extends longer along $y$ (cut 2). The LDOSP (b) is also elongated along $y$, again in the direction of $\v V$. While the ZBP is a natural result of the odd-parity pairing, nodal or nodeless, the elongation of the LDOSP along $\v V$, the direction of gap-maxima in momentum space, is counter-intuitive in a first sight. But this is in fact correct and can be understood as follows. In a quasi-classical point of view, the low energy states slightly away from the vortex core, where the order parameter does not vanish, should have come from super-position of the states near the gap-minima, say $\pm \v k_m$. Since in real space the wave-front of a state $\v k$ is orthogonal to $\v k$, the super-imposed states near $\pm \v k_m$ must be elongated in a real-space direction orthogonal to $\pm \v k_m$. In our case, this elongation is exactly along the $d$-vector of the nematic triplet. The same picture applies to the case of $d$-wave pairing in cuprates. In this case we have two orthogonal nodal directions, and the wave-fronts of quasiparticles from the two pairs of nodes are orthogonal, leading to X-shaped elongation of the LDOSP around a vortex.\cite{Zhu} A further example is the elongation of LDOSP along $x$ for a $p_x$-wave vortex (see SM).

{\em Summary and discussion}: We demonstrated that it is possible to visualize the $d$-vector directly by STM. At subgap energies the $d$-vector (or the direction of gap maxima) is along the leading peak in QPI for a potential impurity, and is along the elongation of the LDOSP of the vortex. 

Since the $D_{4y}$ pairing is nodeless in the presence of warping while $D_{4x}$ is always nodal, the measurement of the bulk DOS would be able to distinguish $D_{4x}$ versus $D_{4y}$. However, if the minimal gap is small in the case of $D_{4y}$, it would be difficult to do the job using bulk DOS alone. The QPI and vortex elongation discussed here are not sensitive to whether the minimal gap is exactly zero and may therefore be most useful. With sufficient resolution, the QPI is most straightforward to pin down the $d$-vector. Interestingly, vortex elongation is indeed observed in recent STM measurements on Bi$_2$Te$_3$/FeTe$_x$Se$_{1-x}$ \cite{Wen} and in Cu$_x$Bi$_2$Se$_3$.\cite{Feng}. High-resolution QPI is yet to be achieved experimentally. Agreement between QPI and vortex elongation regarding the $d$-vector should provide a stringent test of the pairing symmetry. \\

\acknowledgements{WCB thanks Zhao-Long Gu, Da Wang, Lin Yang, and Ming-Hong Jiang for helpful discussions. 
QHW is supported by National Key Research and Development Program of China (under grant No. 2016YFA0300401) and NSFC (under grant No. 11574134). WCB is supported by Zhejiang Provincial Natural Science Foundation of China (Grand No. LQ18A040002).}

\end{document}